\documentclass[aps,prb,twocolumn,superscriptaddress,showpacs,floatfix]{revtex4}
\usepackage{amsmath}
\usepackage{graphicx}
\usepackage{amssymb}
\usepackage[usenames]{color}
\usepackage{ulem}
\usepackage{soul}
\setstcolor{red}

\begin{document}
\title{Effect of pulse error accumulation on
dynamical decoupling of the electron spins of phosphorus donors in silicon}

\author{Zhi-Hui Wang}
\affiliation{Ames Laboratory, Iowa State University, Ames, IA, 50011, USA}
\author{Wenxian Zhang}
\affiliation{Department of Optical Science and Engineering, Fudan University, Shanghai, 200433, China}
\author{A. M. Tyryshkin}
\affiliation{Princeton University, Princeton, NJ, 08544, USA}
\author{S. A. Lyon}
\affiliation{Princeton University, Princeton, NJ, 08544, USA}
\author{J. W. Ager}
\affiliation{Lawrence Berkeley National Laboratory, Berkeley, CA, 94720, USA}
\author{E. E. Haller}
\affiliation{Lawrence Berkeley National Laboratory, Berkeley, CA, 94720, USA}
\affiliation{University of California at Berkeley, Berkeley, CA, 97420, USA}
\author{V. V. Dobrovitski}
\affiliation{Ames Laboratory, Iowa State University, Ames, IA, 50011, USA}
\date{\today}

\begin{abstract}
Dynamical decoupling (DD) is an efficient tool for preserving
quantum coherence in solid-state spin systems.
However, the imperfections of real pulses can ruin the performance of long DD sequences.
We investigate the accumulation and compensation of different pulse errors in DD using the
electron spins of phosphorus donors in silicon as a test system.
We study periodic DD sequences (PDD) based on spin rotations about two perpendicular axes,
and their concatenated and symmetrized versions.
We show that pulse errors may quickly destroy
some spin states, but maintain other
states with high fidelity over long times.
Pulse sequences based on spin rotations about $x$ and $y$ axes
outperform those based on $x$ and $z$ axes due to the accumulation of pulse errors.
Concatenation provides an efficient way to suppress the impact of pulse
errors, and can maintain high fidelity for all spin components:
pulse errors do not accumulate (to first order) as the concatenation level increases,
despite the exponential increase in the number of pulses.
Our theoretical model
gives a clear qualitative picture of the error accumulation,
and produces results in quantitative agreement with the experiments.
\end{abstract}


\maketitle

\section{Introduction}

The state of a closed quantum system can be represented as
a coherent superposition of the basis states, where the phases
and the amplitudes of the superposition coefficients specify
the state. Coherent superpositions of two-level quantum systems
(qubits) can be cleverly employed for high resolution nuclear magnetic resonance
(NMR) and electronic spin resonance (ESR),\cite{Jones09,Cappellaro05}
highly sensitive magnetometers,\cite{Chernobrod05,Taylor08,Degen08,Balasubramanian08}
quantum information processing and quantum computation.\cite{Loss98,Childress06}
However, any real quantum system interacts with its environment.
The interaction leads to the decay of the coherence between the
system's basis states, and the decohered qubit loses its useful properties.
Different ways have been developed to mitigate decoherence.\cite{Zanardi97,Lidar98,Shor95,Knill00}
Among others, the dynamical decoupling (DD) approach looks extremely promising.
DD is based on the ideas that underlie the spin echo effect.\cite{Hahn50,Slichter}
A specially designed sequence of pulses is applied
to the system, and modifies the system-environment coupling
in such a way that the impact of the environment is averaged out.\cite{Viola98,Viola99}
A large number of efficient DD sequences have been
devised in the context of high-resolution NMR.\cite{Haeberlen}
Experimental techniques for producing the pulses
are well developed, especially in the areas of NMR and ESR.
Furthermore, the threshold requirements for application of DD are very modest.
All these advances make DD an appealing strategy in decoherence suppression, and
DD is actively studied now,
both theoretically \cite{Viola99,Khodjasteh05,RBLiu07,Zhang07,Uhrig07,Cywinski08,Uhrig08,Zhang08,West10}
and experimentally.\cite{Beavan05,Morton08,Ladd08,Uys09,Biercuk09,Du09,Dobrovitski10,Naydenov10,Suter10,deLange10,Ryan10,Bluhm10}

Among different DD schemes, in this paper we focus on
DD based on periodic structure, which has been broadly studied and implemented.
A basic DD scheme is the periodic dynamical decoupling (PDD),
in which pulses are repeatedly applied to the system
with equal inter-pulse delay.\cite{Viola98,Viola99}
To suppress the effect of system-environment interaction more efficiently,
symmetrized versions of periodic dynamical decoupling,\cite{Haeberlen,ViolaEDD}
and the concatenated dynamical decoupling (CDD) have been proposed.\cite{Khodjasteh05,Zhang07,Zhang08,Khodjasteh07}
Performance of DD schemes can be different depending on the noise spectrum of the bath,
and has been extensively studied for different qubit systems.\cite{Cywinski08,Biercuk09,Du09,Pasini10,deLange10,Ryan10,Suter10}

Besides the bath dynamics,
real qubits are subject also to errors in the control pulses, which are often systematic, being caused by the instrumental imperfections.
Many research efforts have been devoted to achieving single pulses with small errors,
or alleviating the effect of finite pulse duration,
including composite pulses, soft pulses and Eulerian DD.\cite{GersteinDybowski,Vaughan72,Levitt,Pryadko05,Pasini08,ViolaEDD}
Yet implementations of DD in QIP involve a large number of pulses
applied over long times, so that even small pulse imperfections could accumulate
and seriously affect the decoupling fidelity.
A comprehensive understanding of how the systematic pulse errors accumulate
in different DD sequences is highly desirable.
The effect of pulse errors induced by the finite pulse duration
has been studied in detail. \cite{Khodjasteh07,UhrigPasiniNJP10,Biercuk09,KhodjastehDCG09}
The effect of systematic errors in the rotation axis and angle on CDD
has been studied theoretically \cite{Khodjasteh05}
and experimentally,\cite{Suter10,Biercuk09PRA}
and has also been discussed for dynamically corrected gates.\cite{KhodjastehDCG09}
Moreover, for many traditional NMR/ESR experiments,
the state to preserve is known (e.g.\ prepared along a certain direction)
and the decoupling sequence can be chosen 
to best maintain this state. The Carr-Purcell-Meiboom-Gill\cite{Slichter}
pulse sequence, where $\pi$-pulses are applied to the system along the $x$-axis,
preserves the spin state along the $x$-axis while the state along the $y$-axis
is destroyed by the accumulation of pulse errors.
Quantum information processing, however,
requires any unknown state of the qubit to be preserved, so
DD must preserve all components of the qubit state.

In this paper we analyze, both theoretically and experimentally,
the performance of several DD protocols in the presence of pulse imperfections,
for different initial states.
As a testbed, we use a macroscopic ensemble of electron spins
of $^{31}$P donors in silicon, which constitute a promising system for
studying many fundamental aspects of QIP.\cite{Kane98,Tyryshkin06,Ladd09}
Instead of errors in a single pulse,
we focus on the accumulation and compensation of systematic pulse errors
on a scale of the whole DD sequence.
We study the performance of two-axis PDD,
in which the $\pi$-pulses are applied along two perpendicular axes alternately,
as well as its symmetrized and concatenated versions.
Decoupling fidelities for spin states along $x$, $y$ and $z$ axes are examined.

We find that, in PDD, certain types of pulse errors accumulated during
the first half period of the sequence are balanced out during the second half,
hence have little affect on the decoupling fidelity.
Sequences with pulse axes along $x$ and $y$ alternately (XY PDD)
are mainly affected by the in-plane component of the errors in pulse axes,
while sequences with pulses along $x$ and $z$ axes (XZ PDD)
are sensitive to multiple types of pulse error.
Such analysis could serve as a guide for choosing
suitable decoupling sequences according to specific experimental situation.

We also find that some spin states are quickly destroyed in the decoupling process,
while other components are maintained with high fidelities over long times.
Such a preservation of a particular spin component does not imply good performance of the decoupling sequence, but oppositely, results from the accumulation of the pulse errors over long times. \cite{KhDV_PRA11}
E.g., XY PDD preserves the spin component along the $z$-axis,
while XZ PDD preserves the $y$ component.
XY-based DD sequences are found to outperform the XZ-based sequences.
We analyze the dependence of DD sequences on the accumulation of pulse errors,
and explain all these effects both qualitatively, with a simplified analytical model
and quantitatively, using numerical simulations.

We also find that the concatenated DD protocols (CDD) with two-axis control
exhibit excellent immunity to pulse errors, and preserve all spin components.
The robustness of CDD against systematic pulse errors
has been shown theoretically earlier.\cite{Khodjasteh05}
Our results provide experimental confirmation of this feature of CDD: in the evolution operator the pulse errors do not accumulate to first order, in spite of
the exponential growth of the number of pulses.

The rest of the paper is organized as follows.
In Sec.~\ref{sec:II}, we introduce the system and the DD sequences to be studied.
We present the theoretical model, and describe the experiments.
In Sec.~\ref{sec:main_results}, we present the results of analytical studies and
numerical simulations, and compare them with the experimental data.
Conclusions are given in Sec.~\ref{sec:conclusions}.

\section{The system and the DD protocols}
\label{sec:II}
\subsection{The phosphorus doped silicon system}
\label{sec:IIA}
Electron spins of P donors in silicon show long relaxation times\cite{Honig56,Feher59}
and coherence times,\cite{Tyryshkin03}
and therefore constitute promising candidates for QIP applications\cite{Kane98,Tyryshkin06,Stegner06,Hollenberg10}
and for studying the basic problems of DD.\cite{Tyryshkin06,Morton08}
Advanced ESR techniques can be used to manipulate the spin states of the P donors.\cite{Tyryshkin03,Morton05b}
In our experiments, isotopically purified bulk silicon samples were used,\cite{Ager05}
with $^{29}$Si concentration $\sim800~$ppm,
and the doping density of phosphorus $\sim5\times 10^{14}$~cm$^{-3}$.
Experiments were performed at a static (quantizing) field of $3500~$Gauss,
and a temperature of $8~$K.
The longitudinal relaxation time of the electron spin of P is
$T_1=25~\text{ms}$ limited by the two-phonon Orbach process.\cite{Tyryshkin03,Orbach61}
The transverse relaxation time determined
by spin echo experiments is $T_2=4.6~$ms,
limited by instantaneous diffusion.\cite{KlauderAnderson,Mims68,Tyryshkin03}
The ``true" T$_2$ of isolated donors extrapolates to about $60~\text{ms}$,
e.g. after suppressing the instantaneous diffusion.\cite{Tyryshkin03}

Dynamical decoupling works well when the inter-pulse delay between the pulses
is short compared
to the typical time scale of the noise of the bath.
In our experiments, the inter-pulse delay time is $\tau=11~\mu\text s$.
We now examine the time scales of possible interactions with the P electron spins.

At donor density $\sim5\times 10^{14}$~cm$^{-3}$,
the typical mutual flip-flop time of the P electron spins due to the dipolar interaction
is of the order of $100~$ms, which is much longer than the inter-pulse delay in the DD
and the total duration of the experiment.
We therefore can regard the P electron spins as independent of each other.

The hyperfine coupling between the donor's electron spin
and the donor's $^{31}$P nuclear spin  is rather large ($\sim 100$~MHz),\cite{Fletcher}
and only one hyperfine line is excited in our experiments
(i.e.\ we are working only within the subspace with a fixed $z$-projection of the
$^{31}$P nuclear spin).
Thus, the presence of the $^{31}$P nuclear spin only slightly renormalizes the
resonance frequency of the donor's electron spin, and can be neglected.

Spectral diffusion induced by the nuclear spins
is a major source of decoherence for Si:P system and other dopants in semiconductors.\cite{deSousa03,Witzel05,Witzel06,Cywinski09}
In Si:P system, the P donors are coupled to their surrounding $^{29}$Si
nuclear spins mainly via contact hyperfine interaction.
The anisotropic corrections caused by the admixture of the
$p$-states and the dipolar contribution to the hyperfine coupling, are small.\cite{HuESEEM}
The dynamics of the $^{29}$Si nuclear spin bath,
either due to the flip-flops induced by the intra-bath dipolar interaction
or due to the electron-mediated virtual spin flips,
could decohere the P spins.\cite{deSousa03,Witzel05,Witzel06,Saikin07,Cywinski09}
However, in our experiment the concentration of $^{29}$Si is low ($\sim$800 ppm)
and the quantizing field $B_0=3500~$Gauss, which we take as directed along the $z$-axis, is huge
compared to the hyperfine coupling energy scale.
Both types of flip-flops of the nuclear spins are thus greatly suppressed on the time scale of $\tau$
and on the timescale of the total experiment.
The characteristic spectral diffusion time has been measured to be about $20~$ms
at 800 ppm of $^{29}$Si,\cite{Abe10}
which is larger than the inter-pulse delay by four orders of magnitude.
Therefore the $^{29}$Si nuclear spins can be treated as static,
contributing only an extra static field acting upon the P spins along the $z$-axis.
Moreover, since the separation between the donor centers
is very large, the $^{29}$Si nuclei which are sufficiently strongly coupled to one
center interact very weakly with other donor centers and
each P electron spin can be considered as coupled to its own nuclear spin bath.

The quantizing magnetic field of the magnet also fluctuates.
The fluctuations have a broad spectrum but the noise power quickly decays as $1/f^2$
to become negligible at frequencies higher than $1~$kHz.
The noise correlation time is therefore much larger than $\tau$ and
we can treat the quantizing field as static but inhomogeneous over the sample.

Therefore, the system under study is an ensemble of independent electron spins.
Each spin feels a different static quantizing field, and an extra static field
contributed by the hyperfine interaction with the $^{29}$Si nuclear spins.
The Hamiltonian describing a single phosphorus electron spin $S$ is then
\begin{equation}
\label{eq:haminit}
H= \omega_e S^z + \gamma_e B_{\rm n} S^z
\end{equation}
where $\omega_e$ is the Zeeman frequency of the electron spin in the static quantizing field,
$\gamma_e$ is the gyromagnetic ratio for the P electron,
and $B_{\rm n}$ is the overall effective field acting on $S$ due to interaction with the bath.
Here and below we adopt the system of units with $\hbar=1$.
Taking into account the spatial inhomogeneity
of the static quantizing field,
we represent $\omega_e=\omega_{e0}+\delta\omega_e$,
where $\omega_{e0}$ is the average over the sample,
i.e.\ the frequency corresponding to the
center of the ESR line.
By performing the standard
rotating-frame transformation,\cite{Slichter,Abragam}
$\omega_{e0}$ is eliminated
and the Hamiltonian (\ref{eq:haminit}) is transformed into
\begin{equation}
\label{eq:ham}
H= \gamma_e BS^z,
\end{equation}
where $B=\delta\omega_e/\gamma_e + B_{\rm n}$
is the total field acting upon $S$ in the rotating frame.
Accordingly, all analysis below is performed in the rotating frame.
The distribution function of this field is determined in part by the initial
density matrix of the bath, and in part by the inhomogeneity of $\omega_e$ over the sample.
The statistical properties of the field $B$ determine
the ESR lineshape.
{Our measurements of the free induction decay show that
the ESR line has a Gaussian shape, hence}
\begin{equation}
\label{eq:GaussianB}
P(B) = \frac{1}{\sqrt{2\pi b^2}}\exp{[-B^2/(2b^2)]}
\end{equation}
{with variance $b=50$~mG, which includes both the rms coupling between the spins of the P donors and the bath, and the inhomogeneity of the external quantizing field.}
This experimental fact will be the starting point of the further analysis.

To sum up, on the time scale of the inter-pulse delay of DD (tens of microseconds),
the electron spins effectively live in a static inhomogeneous background magnetic field.
The dephasing time $T_2^*\sim 1.6~\mu$s of the P spins due to such inhomogeneity
is much smaller than $T_2$.
Sequences of microwave-frequency pulses are applied
to decouple the P electron spins from the overall background magnetic field noise,
and refocus the dephasing. The electron spin state is expected to
be preserved by DD against the dephasing on a time scale of the order of $T_2$,
where instantaneous diffusion becomes important.

\subsection{Experimental setup}
\label{sec:exp}
Our experiments were performed on a Bruker Elexsys 580 spectrometer with specially modified software to allow for generating large numbers (over 1000 in some experiments) of microwave pulses,
and a 20 watt continuous wave solid state microwave power amplifier (Amplifier Research) which maintained microwave phase stability over the long pulse sequences. Experiments were performed at an X-band microwave frequency of $9.8$~GHz and a static (quantizing) magnetic field of $3500~$Gauss, at temperature $8$~K. The typical duration of a $\pi$-pulse was $0.18~\mu\text{s}$. In most of the experiments the delay between refocusing pulses was set to $\tau=11~\mu$s, short compared to the characteristic times of all sources of noise known in our system (see the detailed discussion in Section~\ref{sec:IIA}).


For the donors in our Si sample the measured echo signal decays showed an excessive ``phase noise" developing at longer times (for example, at $\tau > 1$~ms in a Hahn echo experiment) arising from the fluctuating magnetic field $B_0$ of the magnet.\cite{Tyryshkin06} The power spectrum of the field noise (Gauss$^2\cdot {\rm Hz}^{-1}$) varies approximately as $1/f^2$ with an amplitude $\sim 0.5$~mG$\cdot\rm Hz^{-1/2}$ at 10 Hz. In the experiments we used a large Si sample, with enough donors to enable the acquisition of an echo without signal averaging. Thus the field noise could be eliminated with magnitude detection, e.g. squaring and adding the in-phase and quadrature components of the echo signal.\cite{Tyryshkin06} This magnitude detection approach was used for an accurate extraction of the $T_2$ relaxation times from the measured decays.

The large Si sample also means that the microwave magnetic field ($B_p$) is not homogeneous over its entire volume in the pulsed ESR resonator (a Bruker dielectrically-loaded cylindrical cavity 4118X-MD5), leading to a systematic rotation angle error. The extent of $B_p$ inhomogeneity and the resulting rotation errors in our spectrometer setup have been characterized in our previous publications.\cite{Morton05a,Morton05b} The $B_p$ variation over the sample volume can be as large as 10--20\% depending on the sample dimensions.
On the other hand, the rotation axis errors
(e.g. relative phase errors between  $\pi_{\text X}$ and  $\pi_{\text Y}$ pulses)
can be reduced to a sub-degree level
{
using a calibration technique based on phase error amplification (SPAM),\cite{Morton05a}
in which the phase errors accumulate in a way which is similar to how the flip angle errors accumulate in the CP sequences,
and hence the orthogonality of the rotation axes in two pulse-forming channels can be measured with high precision.
}

\subsection{Dynamical decoupling protocols}
In a spin echo experiment, the in-plane magnetization of a spin ensemble
decays due to the precession of the spins under an inhomogeneous magnetic field,
and this dephasing can be refocused
by applying a $\pi$-pulse to all the spins
midway during the evolution.
Pulse sequences like Carr-Purcell-Meiboom-Gill (CPMG)
refocus the dephased ensemble magnetization
by applying a train of equally spaced pulses, each rotating the spin by $\pi$
around the $x$-axis of the rotating coordinate frame. Below we denote such
pulses as $\pi_{\text X}$, and other pulses are denoted similarly (e.g.\
a $(\pi/2)_\text{Y}$-pulse is the pulse rotating the spins by $\pi/2$ around the $y$-axis).
However, the single-axis CPMG protocol has an important drawback. When the
electron spins are prepared along the $y$-axis (by applying a
preparatory $(\pi/2)_\text{X}$ pulse),
the small errors inevitably present in real DD pulses
accumulate very quickly in the course of the decoupling experiment,
and destroy the performance of the decoupling protocol,
like in the original Carr-Purcell pulse sequence.
This problem is absent for CPMG in case of the initial states along $x$-axis,
but the goal of DD is to preserve all components of the spin.
Therefore, more general DD protocols must be considered.
A large number of decoupling sequences have been developed,
{
aiming at maintaining the state of the central spin $S$.
Design and analysis of DD schemes are often based on the Magnus expansion (ME),
which is a cumulant expansion of the evolution operator of the whole system (the qubit and the bath).
DD intervenes the evolution and eliminates the unwanted system-bath couplings in
expansion terms of the ME.
}
Besides the DD sequences based on a periodic pulse structure like PDD\cite{Viola99} and
its concatenated version CDD,\cite{Khodjasteh05}
aperiodic structure like UDD,\cite{Uhrig07} QDD,\cite{West10} etc. were developed recently.
In this paper we focus on the decoupling sequences with periodic structure.

One of the simplest protocols is the periodic dynamical decoupling (PDD),\cite{Viola99} where
the central spin is subjected to
a train of equidistant $\pi$-pulses applied along different axes.
The unitary operators, which
describe the control pulses acting on the central spin, are taken from a group cyclically,
starting from the identity element $I$.
A typical sequence based on the group ${\cal G}=\{I,\sigma^x,\sigma^y,\sigma^z\}$
where $\sigma^x,~\sigma^y$ and $\sigma^z$ are Pauli matrices
has a period
{\tt (I-d-I)($\pi_{\text X}$-d-$\pi_{\text X}$)($\pi_{\text Z}$-d-$\pi_{\text Z}$)($\pi_{\text Y}$-d-$\pi_{\text Y}$)}.
Here $\tt d$ indicates the delay between pulses with time duration $\tau$.
By virtue of the operator algebra of the Pauli matrices, the period
can be reduced to
\begin{equation}
\label{eq:XYPDD}
{\tt d-\pi_X-d-\pi_Y-d-\pi_X-d-\pi_Y}\;,
\end{equation}
This two-axis (XY) based decoupling scheme is termed XY PDD.
When the inter-pulse delay time $\tau$ is short
compared to the inverse of the cutoff frequency, $\omega_c$,
of the spectral density function of the bath,
the resulting evolution operator of the system
is equivalent to the identity up to the first order in $\omega_c \tau$, i.e.\ has a form
{
$\mathbf{1}+O(\omega_c^2\tau^2)$.
}
In this paper we also study an alternative two-axis decoupling sequence, XZ PDD.
Its period
\begin{equation}
\label{eq:XZPDD}
{\tt d-\pi_X-d-\pi_Z-d-\pi_X-d-\pi_Z}\;.
\end{equation}
can also be drawn from the group $\cal G$.
For ideal pulses (e.g., when the bath is completely static and pulse errors are absent),
XY- and XZ-based DD are equivalent to each other.

SDD is a symmetrized version of PDD.
The period of SDD sequence is twice {as long as} that of PDD
and is symmetric with respect to the middle.\cite{Haeberlen,Viola99,ViolaEDD}
The period for a XY-based SDD is
\begin{eqnarray}
\label{eq:XY8}
& &{\tt d-\pi_X-d-\pi_Y-d-\pi_X-d-\pi_Y}\nonumber\\
& &{\tt -\pi_Y-d-\pi_X-d-\pi_Y-d-\pi_X-d}\;.
\end{eqnarray}
{Such symmetrization eliminates all even terms in the ME systematically,
leaving the system-bath coupling in the third and higher odd order terms.
}

A concatenated version of periodic decoupling, CDD,
was developed to decouple the system from the bath to higher orders in ME.\cite{Khodjasteh05}
CDD at level one is just PDD,
and sequences for higher levels are built up by
recursively nesting the lower level sequence within itself.
E.g., for the XY-based sequence, the structure for $n$-th level CDD, $\tt CDD_n$, is
\begin{eqnarray}
{\tt CDD_{n-1}{-\tt \pi_X-}CDD_{n-1}{-\tt \pi_Y-}CDD_{n-1}{\tt-\pi_X-}CDD_{n-1}{\tt-\pi_Y}}
\end{eqnarray}
In ME,
$\tt CDD_n$ eliminates
the interaction between the system and the bath
up to the $n$-th order.
Note here the number of pulses in CDD increases approximately as $4^n$.

As is customary in NMR/ESR experiments,
the initial state of the electron spin is pseudo-pure.\cite{Slichter,Abragam}
The density matrix for the electron spin polarized along the $z$-axis
in the high-temperature approximation has the form
$\rho_{0}=\frac{1}{2}{\mathbf 1}+\zeta S^z$ where $\zeta$ is small.
Since the identity matrix ${\mathbf 1}$ is not affected by unitary evolution
and gives no contribution to the signal,
we can regard the electron spin as being in a pure state with $S^z=1/2$
(pseudo-pure state).\cite{Slichter,Abragam}
The in-plane initial states of the electron spin prepared
by $(\pi/2)$-pulses at the beginning of the experiment
then can also be considered as pure.

For a single electron spin $S$,
we characterize the performance of a given DD protocol
using the survival probability of the state (input-output fidelity), which in our case can be written in the form
{
$
\text{Tr}[\rho(0)\rho(t)],
$
where $\rho(0)$ is the initial state of the electron spin
and $\rho(t)$ is the reduced density matrix of the electron spin at the end of DD.
By virtue of the relation $\rho(t)=\mathbf 1/2+\langle S^x\rangle \sigma_x
+\langle S^y\rangle \sigma_y+\langle S^z\rangle \sigma_z$,
the fidelity is equivalent to
\begin{eqnarray}
F^S_\alpha(t)=2\text{Tr}[S^\alpha(t)S^\alpha(0)]=2\text{Tr}[\rho_\alpha(t)S^\alpha]\;,
\end{eqnarray}
and is characterized by the average spin projections $2\langle S^x\rangle$, $2\langle S^y\rangle$, and $2\langle S^z\rangle$.
}
Here $\alpha=x,y,z$ denotes the three initial states of the electron spin
oriented along the axes $x,~y$ and $z$ respectively,
and $\rho_\alpha(t)$ is the corresponding density matrix of the spin at time $t$.
The fidelity characterizes how well
the initial state of the electron spin is protected by the DD sequence.
$F^S_\alpha(t)$ is then averaged over the ensemble as
\begin{eqnarray}
F_\alpha(t)=\langle F^S_\alpha(t)\rangle\;,
\end{eqnarray}
where $\langle\cdot\rangle$ denotes the ensemble average.
The quantity $F_\alpha(t)$ is the measure of the performance of DD in our study.
Of course, this fidelity may, and in fact does, strongly depend on
the specific initial state.

\subsection{Analysis and model of pulse errors}
\label{sec:err}
In ESR experiments pulse errors can be greatly reduced, e.g.\ by using composite pulses.\cite{Levitt}
Still, in many experiments pulse errors remain an important (and sometimes a major) factor
which limit the performance of the decoupling sequence.
We now undertake a detailed analysis of the possible pulse errors in our experiment,
and arrive at a model to account for the error effect.

Since the duration of a $\pi$-pulse, $0.18~\mu\text{s}$, is
small compared to the inverse ESR linewidth ($\sim T_2^*$),
we treat the pulses as instantaneous unitary rotations.
The validity of this approximation has been confirmed numerically,
by modeling the influence of the dephasing field $B$ (Eq.~(\ref{eq:ham}))
during the pulses.
Taking into account the errors in both the rotation axis
and the rotation angle, the operators for the $\pi$-pulses have forms
\begin{eqnarray}
\label{eq:rotationXY}
U_{\text X}&=&\exp{[-i(\pi + \epsilon_x)({\bf S}\cdot\vec {\bf n})]}\\ \nonumber
U_{\text Y}&=&\exp{[-i(\pi + \epsilon_y)({\bf S}\cdot\vec {\bf m})]}\;,
\end{eqnarray}
where $\vec {\bf n}=(\sqrt{1-n^2_y-n^2_z},n_y,n_z)$
is the actual rotation axis for a nominal $\pi_{\text X}$-pulse,
and $\vec {\bf m}=(m_x,\sqrt{1-m^2_x-m^2_z},m_z)$
is the actual rotation axis for a nominal $\pi_{\text Y}$-pulse.
Small parameters $n_y,~n_z,~m_x,~m_z$ characterize
the error in the rotation axes, and $\epsilon_x$ and $\epsilon_y$
characterize the error in the rotation angles.

\subsubsection{Error sources}
\label{sec:err_source}
An ideal $\pi$-pulse generated by a pulse field
in exact resonance with the Larmor frequency of the central spin
realizes a rotation of the central spin by an angle $\pi$.
In the rotating frame, the pulse field $B_p$ has a step-like form in time,
i.e., the value of the field is zero before and after the pulse,
and constant during the pulse. The duration $t_p$ of the pulse is
determined by $t_p\gamma_e B_p=\pi$.
In this study, we consider the following four sources of pulse error.

Due to the inhomogeneity in the static magnetic field over the sample,
the pulse field is not in exact resonance for every electron spin.
In the coordinate frame rotating at angular frequency $\omega_{e0}$,
the detuning, $B$, manifests itself as a non-zero
magnetic field along the $z$-axis.
As a result, the rotation axis deviates from the intended $x$ or $y$ directions,
and the rotation angle $t_p\gamma_e\sqrt{B^2_p+B^2}$ is
different from $\pi$.

In our experiment, the coherence signal was obtained by a single-shot measurement from a
macroscopically large Si sample.
The byproduct of a large sample is that
$B_p$ is not homogeneous over the entire sample volume. 
To model this effect in a simple but physically meaningful way, we assume
a one-dimensional model, where
$B_p$ varies only along some axis $l$ (e.g.\, along the resonator axis),
and the sample position is optimized near the maximum of the driving field, where $B_p$
depends quadratically on $l$:
\begin{equation}
\label{eq:BpX}
B_p(l)= \bar B_{p} +\Delta B_p[1-3l^2/d^2]\;,
\end{equation}
where $\bar B_p$ is the average of $B_p(l)$ over the sample,
and $\Delta B_p$ quantifies the amplitude of the error.
The origin of the $l$-axis is at the center of the sample,
and $2d$ is the sample length along the $l$-axis.
In the experiment, $B_p$ can be well tuned such that
on average an exact $\pi$-pulse can be expected, i.e.\
$t_p\gamma_e\bar B_p =\pi$.
The error in the rotation angle $\pi+\epsilon(l)$
for the spin located at position $l$
then arises from the $\Delta B_p$ term
\begin{equation}
\epsilon(l)=t_p\gamma_e\Delta B_p (1-3l^2/d^2)\;.
\end{equation}
The corresponding distribution function for the rotation angle error is
\begin{equation}
\label{epsdist}
P(\epsilon)=(1/2\epsilon_0)[3(1-\epsilon/\epsilon_0)]^{-1/2}\;.
\end{equation}
and $-2\epsilon_0\leq\epsilon\leq \epsilon_0$.
We further assume the angle errors for $\pi_{\text X}$- and $\pi_{\text Y}$- pulses to be the same $\epsilon_x=\epsilon_y=\epsilon$.

The imperfectly rectangular shape of the actual pulses also introduces pulse errors.
These errors mainly come from the leading and trailing edges of the pulse.
In ESR experiments, 
the edges may constitute about 10\% of the total pulse time,
and could have a noticeable influence.
At the pulse edges, the amplitude and the phase of the driving field are ill-defined,
introducing errors to both the rotation axis and the rotation angle.
Furthermore, the magnitude of these transient errors depends on the offset frequency, $B$,
of the spin and therefore the
pulse errors have a non-uniform distribution over the sample.
To account for the effect of these complex errors,
we introduce the rotation axis errors $n_y,m_x$ and $n_z,m_z$.
Similar to the arguments used in Eq.~(\ref{epsdist}),
the axis error, $n_z$, is drawn from the probability distributions
\begin{eqnarray}
\label{nzdist}
P(n_z)&=&(1/2n_0)[3(1-n_z/n_0)]^{-1/2}\;,
\end{eqnarray}
where $-2n_0\leq n_z\leq n_0$,
and constant $n_0$ quantifies the amplitude of the error.
Similarly, we assume the axis errors of $\pi_{\text Y}$-pulses have the
same probability distribution as Eq.~(\ref{nzdist}),
and assume $m_z=n_z$ for all spins.
The value for the in-plane component of
the rotation axis errors, $m_x$ and $n_y$, are extremely
small in experiments as discussed in Section~\ref{sec:exp},
so we take them to be zero in the simulations,
but keep the symbols for the analysis.

Another possible pulse error is the imperfection
in the relative phases of the $\pi_{\text X}$ and $\pi_{\text Y}$-pulses.
In the presence of such errors, if we assume that
the axis for $\pi_{\text X}$-pulse
is perfectly aligned along the $x$-axis,
the actual rotation axis of the $\pi_{\text Y}$-pulse may have a non-zero $x$-component.
However, by using standard phase calibration techniques,
this in-plane axis error can be reduced to a sub-degree level and is negligible in our experiments.

Summarizing,
we treat the pulses as instantaneous rotations described by Eq.~(\ref{eq:rotationXY}).
The pulse errors
$\epsilon_x=\epsilon_y=\epsilon$ and $n_z=m_z$ are drawn
from distributions Eq.~(\ref{epsdist}) and Eq.~(\ref{nzdist}), respectively.
Note that in our model for pulse errors, $\epsilon_0$ and $n_0$ are the only adjustable parameters
which are determined from experiments.
With a fixed set of values for $\epsilon_0$ and $n_0$,
good simulation results are achieved in quantitative agreement
with the experiments for all seven different DD protocols
(XY and XZ PDD and CDD. Symmetrized DD sequences were not performed experimentally).

\subsubsection{Simulation details}
Time evolution of the electron spin $S$ is simulated for
initial states along axes $x$, $y$ and $z$.
We label these three initial states respectively
as $|\psi_x\rangle$, $|\psi_y\rangle$ and $|\psi_z\rangle$,
and the fidelities as $F_x$, $F_y$ and $F_z$.
Simulations are performed in the rotating frame
with the frequency at the center of the ESR line, $\omega_{e0}$.
For each single spin,
the pulses are implemented as described by Eq.~(\ref{eq:rotationXY}), and
the evolution operator for the inter-pulse delay is
$U_0=\exp{[-i H t]}$ with the Hamiltonian $H$ given by Eq.~(\ref{eq:ham}).
The static field $B$ felt by spin $S$ is drawn from the Gaussian distribution, Eq.~(\ref{eq:GaussianB}).
After application of the decoupling sequence,
the input-output fidelities, $F^S_\alpha(t)$, are calculated.
The fidelities are averaged over $\sim 10^4$ realizations.
The time delay between pulses is constant $\tau=11~\mu$s
for all DD sequences as in the experiments.
The values of the error parameters in Eqs.~(\ref{epsdist}), (\ref{nzdist})
are taken as
$\epsilon_0=0.3~(7.5^\circ)$ and $n_0=-0.12~(-3.5^\circ)$,
which provides results that fit well with the experiments.

\section{Results and Discussion}
\label{sec:main_results}
Experimental and numerical results for different DD protocols
are shown in Figure~\ref{fig:PDD} through Figure~\ref{fig:XY8CDD}.
The simulation results agree well with the experimental data, indicating that
the error model captures the essential features of the system.
We present the results for decoupling sequences based on XY and XZ pulses for comparison.
In the experiment, a $\pi_{\text Z}$ pulse was implemented by a pair of subsequent rotations about
the $x$ and $y$ axes, $\pi_{\text X}\pi_{\text Y}$.
This substitution was carried out in experiments for all the XZ based sequences.
Since the bath of the electron spins is static in our study,
for ideal pulses (without pulse errors),
XY- and XZ- based sequences are exactly equivalent.
However, as we show below, in the presence of pulse errors
they behave very differently
in preserving the quantum states of the electron spin .

\subsection{PDD}
\subsubsection{XY-based sequence}
\label{sec:XYPDD}
\begin{figure}[htbp]
\begin{center}
\includegraphics[width=4in,angle=270]{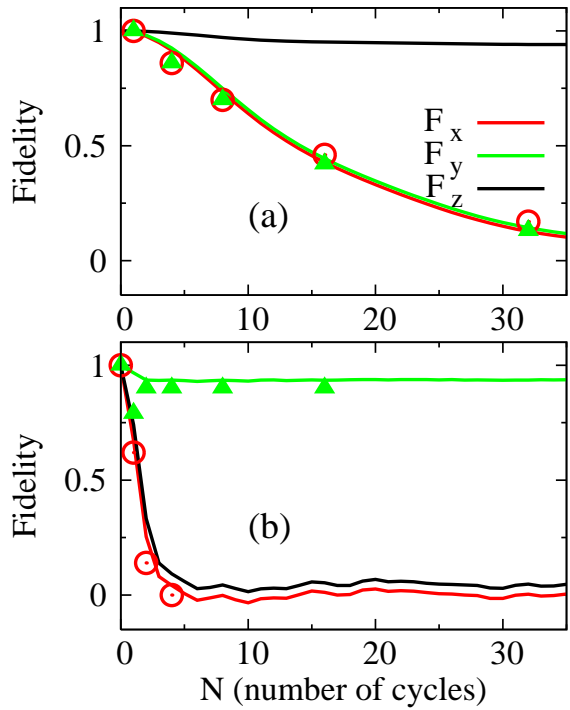}
\end{center}
\caption[]
{\label {fig:PDD}
(color online).
Fidelity as a function of the number of cycles, $N$, of PDD.
(a) XY PDD, (b) XZ PDD for initial states $|\psi_x\rangle$, $|\psi_y\rangle$,
and $|\psi_z\rangle$, respectively.
Points are experimental results and lines are simulation results.
}
\end{figure}

Figure~\ref{fig:PDD} (a) shows
the fidelities of the electron spin states
as functions of the number of periods (cycles) for XY PDD.
Fidelities for different initial states exhibit different decay behaviors.
$F_x$ and $F_y$ exhibit apparent decays.
$F_z$, on the contrary, decays very little.
Similar phenomenon in DD have been discussed in Ref.~[\onlinecite{Zhang07}],
where one state of the system is less sensitive to the decoherence
induced by the spin-bath interactions than the other states.
In our study, in the presence of pulse imperfections,
particular initial states of the electron
are preserved, i.e., survive for a long time
in the DD process,
while other states are quickly destroyed by the accumulation of pulse errors.

The electron spins of P are independent of each other, and are
coupled only to external magnetic fields (static field along the $z$-axis, and
a time-varying pulse fields along all three axes). In this case,
the time evolution of each spin is unitary, and
can be described as a rotation:
\begin{equation}
\label{eq:U}
U=\exp{[-i \theta ( {\bf S} \cdot \vec {\bf a}  )]}\;,
\end{equation}
where $\vec {\bf a}$ is the effective rotation axis
and $\theta$ is the rotation angle.
In the following we will examine such rotations,
viewed stroboscopically after each period of the DD sequence.

%
%

Consider a period of the XY PDD sequence Eq.~(\ref{eq:XYPDD}),
substituting the expressions for pulse errors,
Eq.~(\ref{eq:rotationXY}), into the evolution operator
\begin{eqnarray}
\label{eq:U2}
U^{\tt XYPDD}&=&U_{\text Y}U_0U_{\text X}U_0U_{\text Y}U_0U_{\text X}U_0\;,
\end{eqnarray}
and keeping only the zeroth and first order terms
in the small parameters ($\epsilon_x,~\epsilon_y,~m_x,~n_y,~m_z,~n_z$ ),
$U^{\tt XYPDD}$ can be expressed in form of Eq.~(\ref{eq:U}) with
\begin{eqnarray}
\label{eq:axisXY}
\vec {\bf a}&=&(0,0,-1)\nonumber\\
\theta &=&2\pi+\delta\theta
\end{eqnarray}
where
\begin{equation}
\label{eq:thetaXY}
\delta\theta =4(m_x+n_y)\;.
\end{equation}
That is, after one period,
the spin is rotated about an axis close to $-z$
by an angle $2\pi+\delta\theta$.
For ideal pulses, $U^{\tt XYPDD}=-1$, the rotation angle is $\theta=2\pi$
and the spin returns exactly to its initial state.
The error in the rotation angle, $\delta\theta$,
can be viewed as a dephasing error in the plane perpendicular to $\vec {\bf a}$.
After $N$ periods, to first order in the pulse errors,
the spin rotates about the axis $\vec {\bf a}$ by an angle $N\delta\theta$.
Since $\vec {\bf a}$ is (approximately) along the $z$-axis,
the spin operator $S^z$ (approximately) commutes with the evolution operator, Eq.~(\ref{eq:U}).
Therefore the initial state $|\psi_z\rangle$ is an (approximate) eigenstate of the evolution operator,
and thus does not evolve with the DD sequence, and is of course insensitive to pulse errors
(pulse errors in high order could contribute though).
On the other hand, the other two states $|\psi_x\rangle$ and $|\psi_y\rangle$
are not the eigenstates and for these states the error in the rotation angle
$\delta\theta$ will accumulate during the pulse sequence.

Now we consider the electron spin ensemble.
For each spin the time evolution is governed by a rotation, Eq.~(\ref{eq:U}),
with rotation axis and angle given by Eq.~(\ref{eq:axisXY}).
Each spin sees different pulse errors,
and the resultant effective rotation axis, $\vec {\bf a}$, and dephasing errors, $\delta\theta$,
are slightly different from each other (for first or higher order pulse errors).
As shown in Eq.~(\ref{eq:axisXY}), the rotation axes for all spins are along $z$
up to first order in the pulse errors.
The initial state $|\psi_z\rangle$ is then preserved for all spins in the ensemble,
despite their different pulse errors.
The ensemble-averaged fidelity, $F_z$, hence remains high
until spoiled by the accumulation of higher order errors.
On the other hand, for states $|\psi_x\rangle$ and $|\psi_y\rangle$,
the spins are initially in the plane perpendicular to the $\vec {\bf a}$ ($z$)-axis.
Different spins acquire different phase errors
during the evolution in one period of DD.
As a number of periods $N$ increases, this error accumulates.
After certain number of periods, the spin components spread out evenly in the $x$-$y$ plane,
and the ensemble-averaged fidelity decays to almost zero.

Note here if we analyze the rotation operator after half a period
${\tt d-\pi_X-d-\pi_Y}$, (which is the repeating unit of the pulse sequence),
and express the corresponding evolution operator $U^{\tt XYPDD}_{\tt half}$
up to first order in the pulse errors,
the rotation angle is $\theta'=\pi+2(m_x+n_y)$
and the axis of the rotation, $\vec {\bf a}'$, is
\begin{eqnarray}
a'_{x} & = & -\frac{\epsilon_{y}}{2}+n_{z}\cos \phi_{\text d}-\frac{\epsilon_{x}}{2}\sin \phi_{\text d}\nonumber\\
a'_{y} & = & m_{z}-\frac{\epsilon_{x}}{2}\cos \phi_{\text d}-n_{z}\sin \phi_{\text d}\nonumber\\
a'_{z} & = & -1
\end{eqnarray}
where $\phi_{\text d}=\gamma_e B\tau$ is the phase accumulated during the inter-pulse delay.
It is noteworthy that unlike the rotation axis for a full period in Eq.~(\ref{eq:axisXY}),
the rotation axis $\vec {\bf a}'$ involves pulse errors to first order.
That is, some pulse errors accumulated during the first half period
are balanced out during the second half.
%
Since a full period is a repetition of two half periods, we certainly have
\begin{eqnarray}
U^{\tt XYPDD}=U^{\tt XYPDD}_{\tt half}U^{\tt XYPDD}_{\tt half}=\exp{[-i 2\theta' ( {\bf S} \cdot \vec {\bf a}'  )]}\;,
\end{eqnarray}
and the rotation axis is still the same and the rotation angle is twice larger.
Although the expressions for $\vec {\bf a}$ and $\vec {\bf a}'$ are different, there is no discrepancy here.
Since $\exp{[-i 2\theta' ( {\bf S} \cdot \vec {\bf a}'  )]} = \cos\theta' -i/2 \sin\theta' ({\bf S} \cdot \vec {\bf a}')$
and $\sin\theta'=-2(m_x+n_y)$,
first order pulse errors in $\vec {\bf a}'$ become higher order in $U^{\tt XYPDD}$.
Therefore, up to fist order, $U^{\tt XYPDD}$ can also be expressed as Eq.~(\ref{eq:axisXY})
with the effective axis, $\vec {\bf a}$, free of first order pulse errors.
We next undertake a more quantitative examination of how the decays and the preservation
happen to different spin components.
For a single spin, $S$, the fidelity
for an initial state, $|\psi_\alpha \rangle$ ($\alpha=x,y,z$), after $N$ periods is
\begin{eqnarray}
\label{eq:fidS}
F^S_\alpha (N)&=&2\text{Tr}\left[U^{\tt XYPDD}\rho_S(0)(U^{\tt XYPDD})^\dag S^\alpha\right]\nonumber\\
&=&a^2_\alpha+(1-a^2_\alpha)\cos{(N\delta\theta)}\;.
\end{eqnarray}
That is, the fidelity, if viewed stroboscopically (at the end of each DD period),
exhibits oscillations at a frequency determined by the phase error, $\delta\theta$.
For spin ensemble,
after averaging over $B$ and over the pulse errors
(although the magnetic field $B$ does not show up in Eq.~(\ref{eq:thetaXY}),
it contributes in higher order),
the cosine factor contributes to the decay, and the fidelity at long times
saturates at a value independent of time ($N$)
\begin{equation}
F_\alpha(N) \to \langle a^2_\alpha\rangle\;,
\end{equation}
where $\langle\cdot\rangle$ denotes the ensemble average.
For $|\psi_x\rangle$ and $|\psi_y\rangle$, $\langle a^2_\alpha\rangle$ is small,
hence the fidelities decay almost to zero. For $|\psi_z\rangle$,
since $a_z$ is almost one, the fidelity is preserved with its value close to 1.

Such preservation of a particular spin state
demonstrates a remarkable feature of the error accumulation in DD:
a large fidelity for certain initial states does not guarantee
good performance of DD for other initial states, hence
does not necessarily suggest a good DD protocol.
This is similar to the error accumulation effect in the single-axis DD sequences.
While $|\psi_x\rangle$ is well preserved by the $\pi_{\text X}$-pulse sequence (CPMG),
$|\psi_y\rangle$ is quickly destroyed by the same sequence due to
the accumulation of errors in the rotation angle (CP).
{
Sensitivity of Uhrig DD\cite{Uhrig07} and CPMG to initial input state
in the presence of pulse errors also has recently been studied experimentally in Ref. [\onlinecite{Biercuk09PRA}].
}
Therefore, in experiments implementing DD,
the preservation of all three components $F_x$, $F_y$ and $F_z$
should be checked, or full tomography should be performed.

%

\subsubsection{XZ-based sequence}
The pulse sequence of XZ PDD preserves the spin component $|\psi_y\rangle$, see Figure~\ref{fig:PDD} (b).
Compared to the decaying fidelities ($F_x$ and $F_y$) in XY PDD,
the fidelities $F_x$ and $F_z$ in XZ PDD decay much faster,
with values below 0.2 after barely 3 cycles (12 pulses).

An analysis similar to that in \ref{sec:XYPDD} shows that
the evolution operator for one period of XZ PDD
$U^{\tt XZPDD}=\exp{[-i(2\pi+\delta\theta) ({\bf S} \cdot \vec {\bf a} )]}$,
up to first order in the pulse errors, corresponds to
\begin{eqnarray}
\label{eq:err_XZPDD}
\vec {\bf a}&=&(0,-1,0)\nonumber\\
\delta\theta &=&2[-\epsilon_y+\epsilon_x\sin{\phi_{\text d}}+2n_z(1-\cos{\phi_{\text d}})]\;.
\end{eqnarray}
The effective rotation axis $\vec {\bf a}$ is close to $-y$, hence the state $|\psi_y\rangle$ is preserved.

Comparing Eq.~(\ref{eq:err_XZPDD}) to Eq.~(\ref{eq:axisXY}),
one can see that XZ PDD involves rotation angle errors $\epsilon_x$, $\epsilon_y$
and axis error $n_z$ to first order,
while the XY-based sequence involves only $m_x$ and $n_y$.
Since in experiments, as we mentioned in Section~\ref{sec:err},
the in-plane rotation components, $m_x$ and $n_y$, are extremely small and can be neglected,
the deviation of the evolution operator of a period of XY PDD from identity
is actually second order in the pulse errors
while that of XZ PDD is first order.
Furthermore, note that Eq.~(\ref{eq:err_XZPDD})
for XZ PDD is dependent on the magnetic field (in $\phi_{\text d}$)
while Eq.~(\ref{eq:axisXY}) for XY PDD is not.
When performing the ensemble average in Eq.~(\ref{eq:fidS}), which holds
also for XZ PDD and in general for any periodic decoupling sequence,
the first order pulse error in $\delta\theta$ and the $B$-dependence
leads to a much faster decay of the cosine term
than the second order pulse error does for XY PDD.
This explains why XY PDD outperforms XZ PDD.

Note that such inferior performance of XZ PDD compared to XY PDD
does not arise from the replacement of $\pi_{\text Z}$
by $\pi_{\text X}\pi_{\text Y}$ in our experiment.
The XZ-based sequence still performs worse
even if $\pi_{\text Z}$-pulses are applied directly.
Taking the form of the evolution operator for $\pi_{\text Z}$
to be similar to Eq.~(\ref{eq:rotationXY}),
with rotation axis $\vec {\bf p}$
and rotation angle $\pi+\epsilon_z$,
Eq.~(\ref{eq:err_XZPDD}) becomes
\begin{eqnarray}
\label{eq:XZ2}
\vec {\bf a}&=&(0,-1,0)\nonumber\\
\delta\theta &=&2[-2p_x+\epsilon_x\sin{\phi_{\text d}}-2n_z\cos{\phi_{\text d}}]\;,
\end{eqnarray}
where $p_x$ is the $x$-component of the rotation axis for the $\pi_{\text Z}$-pulse.
(Note, in the limit $\phi_{\text d} \to 0$, we have $\delta\theta = -4(p_x+n_z)$,
which is exactly the symmetric form of their counterpart for the XY-based sequence Eq.~(\ref{eq:axisXY}),
as expected from the rotational symmetry of the system.)
The evolution operator for XZ PDD hence contains multiple first order pulse errors,
similar to the case where  $\pi_{\text Z}$ is replaced by $\pi_{\text X}\pi_{\text Y}$.
Therefore, given the experimental facts that the in-plane component of the pulse axis error,
$m_x$ and $n_y$, can be made extremely small while the other errors cannot be neglected,
we can expect XZ PDD to perform worse than XY PDD.

In summary, pulse errors play an important role in periodic DD.
Each PDD sequence has an effective axis in the presence of pulse errors.
During the decoupling, viewed stroboscopically, the spins perform rotations about this axis.
Error accumulation could destroy the initial state components perpendicular
to the effective rotation axis, while the component parallel to the axis is preserved.
For negligible pulse axis errors, $m_x$ and $n_y$, XY based decoupling sequence
shows a better resistance to the pulse imperfections than the XZ-based sequence.

\subsection{CDD}
Next we examine concatenated sequences based on PDD.
We measured the fidelity for a single cycle of XY and XZ based CDD,
with concatenation levels 1 to 4.
Correspondingly, the numbers of pulses are 4, 20, 84 and 340,
and the number of delays (all delays have the same duration $\tau$)
are 4, 16, 64, and 256, respectively.
Fidelities are evaluated after the application of the whole sequence.

\begin{figure}[htbp]
\begin{center}
\includegraphics[width=4in,angle=270]{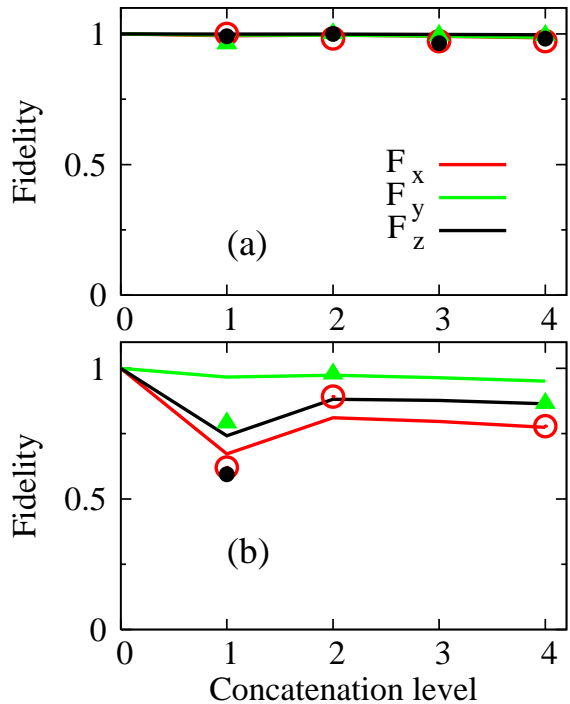}
\end{center}
\caption[]
{\label {fig:CDD}
(color online).
Concatenated DD.
Fidelity as a function of the concatenation level
is shown for initial states $|\psi_x\rangle$, $|\psi_y\rangle$ and $|\psi_z\rangle$.
Points are experimental results and lines are simulation results.
(a) XY CDD. All three lines for states $|\psi_x\rangle$, $|\psi_y\rangle$,
and $|\psi_z\rangle$ overlap.
(b) XZ CDD.
}
\end{figure}

In XY CDD, as shown in Figure~\ref{fig:CDD} (a),
all three initial states are preserved with excellent fidelities ($>99\%$).
Numerical simulations based on our error model are
in agreement with the experimental results.
Note here, at level 4, the electron spin has been subjected to 340 pulses,
and its quantum state is perfectly preserved with fidelity close to unity
for $2.8~$ms, which is of the order of $T_2$.
For many pulses and over a long time,
no visible sign of the error accumulation appears.

To better understand the excellent performance of XY CDD,
we analyze error accumulation in the concatenation.
The evolution operator for a period of the level-2 CDD $U^{\tt XYCDD2}$,
up to first order in the pulse errors,
corresponds to a rotation with
rotation axis and angle
\begin{eqnarray}
\label{eq:err_XYCDD2}
\vec {\bf a}&=&(0,0,-1)\nonumber\\
\delta\theta &=&4(m_x+n_y)\;.
\end{eqnarray}
This exactly repeats the rotation corresponding to XY PDD, Eq.~(\ref{eq:axisXY}).
Further examining higher concatenation levels, we arrive at the following relations
\begin{equation}
\label{eq:repeatXYCDD}
U^{\tt XYPDD}=U^{\tt XYCDD2}=\cdots=U^{\tt XYCDDn}\;,
\end{equation}
where $U^{\tt XYCDDn}$ denotes the evolution operator for level-$n$ concatenation.
Up to first order in the pulse errors,
the evolution operators are exactly the same for all concatenation levels.
Although the number of pulses increases exponentially with the concatenation level,
the first order errors do not accumulate.
Concatenation endows DD
with the capability of self-correcting the pulse errors.


XZ-based concatenated DD also outperforms its PDD counterpart, see Figure~\ref{fig:CDD} (b).
However, compared to XY CDD, the performance is worse.
While $|\psi_y\rangle$ is the saturating component,
the fidelities for the other two states are lower.
Notably, at the first concatenation level
the fidelity is smaller than at the other levels.
Relations similar to Eq.~(\ref{eq:repeatXYCDD}) hold for XZ CDD as
\begin{equation}
\label{eq:repeatXZCDD}
U^{\tt XZPDD}|_{B\tau= 0}=U^{\tt XZCDD2}=U^{\tt XZCDD3}=\cdots=U^{\tt XZCDDn}\;.
\end{equation}
For the first level concatenation (PDD) the evolution operator
depends on the phase during the inter-pulse delay, $\phi_{\text d}$,
i.e.\ depends on the factor $\gamma_e B\tau$, but the higher levels are independent.
This explains the dips at level-1 concatenation for $F_x$ and
$F_z$ in Figure~\ref{fig:CDD} (b).
Therefore, the concatenation eliminates the dependence
of the error accumulation on the inter-pulse delay.

The simulation results for XZ CDD are qualitatively in agreement
with the experimental data, e.g.\ state $|\psi_y\rangle$ is
better preserved than the other two states, and there is a dip
in the first-order concatenation for $|\psi_x\rangle$.
The quantitative difference may be due to other experimental details
that we did not take into consideration in our error model.


\subsection{Symmetrized DD}
\begin{figure}[htbp]
\begin{center}
\includegraphics[width=2in,angle=270]{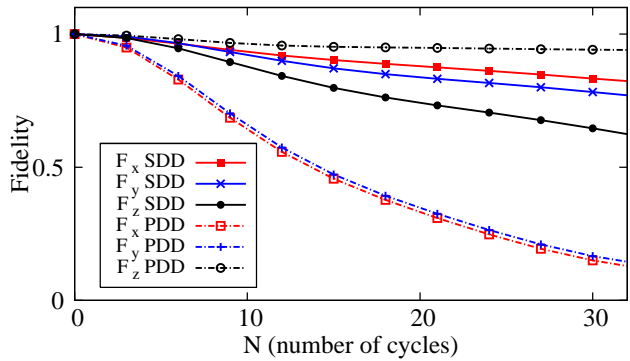}
\end{center}
\caption[]
{\label {fig:XY8PDD}
(color online).
{Simulation results for} XY-based SDD.
Fidelity as a function of the number of cycles is shown
for different initial states.
Fidelities in XY PDD are plotted as broken lines for comparison.
Note that the number of pulses in one cycle of SDD is twice {as large as} that in PDD.}
\end{figure}

We also examined the symmetrized XY-based decoupling sequence
via numerical simulations and theoretical analysis.
The symmetrized pulse sequence unit Eq.~(\ref{eq:XY8}) is repeated periodically (XY-SDD).
Figure~\ref{fig:XY8PDD} shows the simulation results.
The fidelities of the spin states exhibit much slower overall decay
compared to XY PDD.
Indeed, thanks to the symmetrization, the first order pulse errors balance out
in the evolution operator for a single period of SDD.
Up to second order, the evolution operator reads
\begin{eqnarray}
\label{eq:SDD}
U^{\tt XY{\text-}SDD}&=&{\mathbf 1}+2i\epsilon_y(m_x+n_y)\sigma_x\nonumber\\
& &+2i(m_x+n_y)\left(\epsilon_x\cos{\phi_{\text d}}
+2n_z\sin{\phi_{\text d}}\right)\sigma_y\;.\nonumber\\
\end{eqnarray}
Furthermore, since in our case the in-plane axis errors, $m_x$ and $n_y$, are negligible,
pulse errors actually contribute only in the 3rd and higher orders.
The good performance of SDD in Figure~\ref{fig:XY8PDD} compared to XY PDD
is hence expected.
{
The SDD sequence we study here has a similarity in spirit with
the Eulerian decoupling (EDD) \cite{ViolaEDD}.
By making the structure of the decoupling sequence more symmetric,
both SDD and EDD cancel the pulse errors to first order.
}

Recall that in XY PDD, spin component $|\psi_z\rangle$
is preserved against pulse errors for a much longer time
than the other two components.
For SDD, such asymmetric behavior is less obvious.
This is because to 3rd order,
$U^{\tt XY{\text-}SDD}$
is a rotation operator with the rotation axis in the $x$-$y$ plane
(the expression is cumbersome and not shown here).
Symmetrized DD treats the three spin components on a more equal footing.

\begin{figure}[htbp]
\begin{center}
\includegraphics[width=4in,angle=270]{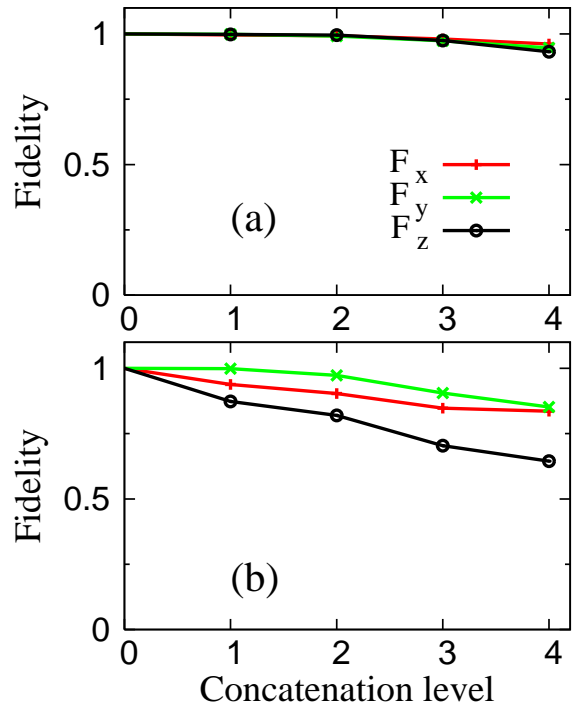}
\end{center}
\caption[]
{\label {fig:XY8CDD}
(color online).
{Simulation results for} concatenated XY-based SDD.
Fidelity as a function of the concatenation level
is shown for different spin states.
(a) Results for the full sequence.
(b) Results for the reduced sequence (with all adjacent $\pi_{\text Y}$-pulses removed).}
\end{figure}

In the symmetrized sequence, Eq.~(\ref{eq:XY8}),
two $\pi_{\text Y}$-pulses are adjacent to each other in the middle of a period.
Since $\sigma^y\sigma^y={\mathbf 1}$, one would expect that removing these two
imperfect pulses from the sequence will yield better fidelity. 
However,
the concatenated decoupling sequence
based on such a reduced version of the SDD behaves much worse
than the full sequence, as seen in Figure~\ref{fig:XY8CDD}.
This can also be understood from the error dependence of the evolution operator.
While $U^{\tt XY{\text-}SDD}$ is identity
up to first order in the pulse errors,
with the two $\pi_{\text Y}$-pulses removed,
the evolution operator involves $\epsilon_y$ to first order
\begin{eqnarray}
U^{\tt XY{\text-}SDD}_{\tt reduced}=-{\mathbf 1}-i\epsilon_y \sigma^y\;.
\end{eqnarray}
where the subscript ``reduced" denotes that all adjacent $\pi_{\text Y}$-pulses are removed from the sequence.
Therefore, these two $\pi_{\text Y}$-pulses help keep a balanced structure for
the symmetrized XY sequence to suppress error accumulation.

%

\section{Conclusions}
\label{sec:conclusions}
We have studied the effects of pulse imperfections on dynamical decoupling.
Using the electron spins of phosphorus donors in silicon as a testbed,
we investigated the performance of different DD protocols.
Experimental results are qualitatively explained by theoretical analysis
and quantitatively reproduced by numerical simulations.

Dynamical decoupling sequences are susceptible to pulse imperfections.
The accumulation of small pulse errors can have appreciable influence on the performance of DD.
In two-axis based PDD, the spin component along a certain direction is preserved
with high fidelity for a long time while the
components in the perpendicular directions decay rapidly.
DD sequences are analyzed in terms of spin rotation,
and each DD sequence has an effective axis in the presence of pulse errors.
During the decoupling, viewed stroboscopically, the spins perform rotations about this axis.
The better preservation of a particular spin component
is attributed to the fact that
this component is parallel to the effective
axis of the DD sequence
and hence is a quasi integral of motion, in spite of pulse errors.

The XY-based decoupling protocols prominently outperform their XZ-based counterparts.
The different performance is explained analytically in terms of pulse errors.
A symmetrized DD sequence suppresses the accumulation of first order pulse errors systematically
and thus performs better than regular PDD.
Interestingly, the accumulation of pulse errors is better suppressed
by keeping the seemingly trivial adjacent identical pulses, $\pi_{\text Y}\pi_{\text Y}$, in SDD.
The symmetrized sequence itself has an error-balance structure in which these two pulses are indispensable.

The concatenated dynamical decoupling sequences are found to be error-resistant.
Although the number of pulses increases exponentially with the concatenation level,
CDD have exactly the same error effects up to first order in the pulse errors.
Concatenation can also alleviate the dependence of the
error accumulation on the inter-pulse delay as well as the inhomogeneity in the
static magnetic field.
XY CDD, with the advantage of both the slow accumulation of pulse errors compared to XZ-based sequences
and the superior error-resistance of the concatenation structure,
demonstrates the ability to store all three spin components for a long time.

{
In a general situation, when the internal bath dynamics is important, the joint action of the fluctuating bath and the imperfect pulses leads to a very complex qubit dynamics. However, in the case where the two effects are small, we can, at least on a qualitative level, consider their contributions separately \cite{GersteinDybowski}. As such, our work, focused on the role of the pulse error accumulation, is complementary to the studies of the bath dynamics and the resulting homogeneous dephasing. By treating the bath field as static, we can single out the effect of the pulse error accumulation, and clarify its important role in the dynamical decoupling experiments.
Moreover, the inhomogeneous static broadening is often important by itself. For instance, in the experiments using single qubits with the projective readout (such as most experiments on quantum dots, the nitrogen-vacancy centers in diamond, superconducting qubits, etc.), where many experimental runs are needed in order to build statistics, the quasistatic shot-to-shot variations lead to inhomogeneous dephasing, and have to be refocused.
}


\acknowledgements

We thank D. G. Cory, L. Cywinski, L. Viola and D. Lidar for useful discussions.
We thank E. Hahn for pointing out the possible importance of radiation damping.
We thank R. Weber and P. Hofer of Bruker Biospin for support with instrumentation.
Work at Ames Laboratory (theory and simulations) was supported by the Department of Energy --
Basic Energy Sciences under Contract No DE-AC02-07CH11358.
Work at Princeton was supported by the NSF through the Princeton MRSEC under Grant No. DMR-0213706 and by the NAS/LPS through LBNL under MOD 713106A. Work at the LBNL (Si single crystal synthesis and processing) was supported by the Director, Office of Science, Office of Basic Energy Sciences, Materials Sciences and Engineering Division of the US Department of Energy (DE-AC02-05CH11231).

\appendix
\section{Radiation Damping}
\label{sec:RD}

In NMR and ESR experiments, when the in-plane magnetization signal is strong,
its precession about the static field
induces a noticeable time-dependent current in the resonator.
This current generates an ac magnetic field,
which exerts a torque on the spins,
and causes them to relax back to their equilibrium states.
This phenomenon is called radiation damping (RD).\cite{Bloembergen54,Abragam}
We now consider the effect of RD on dynamical decoupling in our ESR experiments.

The magnetic field generated by RD (the RD field), ${\bf B}_r$,
is in the $x$-$y$ plane, with its magnitude proportional to the in-plane magnetization, ${\bf M}$,
and its direction perpendicular to ${\bf M}$,\cite{Abragam}
\begin{equation}
\label{eq:Br}
\gamma_e{\bf B}_r = \frac{1}{\tau_r} (-{\cal M}_y\hat x + {\cal M}_x\hat y)\;,
\end{equation}
where $\vec{\cal M}$ is the overall magnetization normalized by the equilibrium magnetization, $M_0$
$$
\vec{\cal M}={\bf M}/M_0\;,
$$
and the damping time constant
\begin{equation}
\label{eq:tauR}
\tau_r=\frac{1}{2\pi \gamma_e M_0 Q \eta}
\end{equation}
is determined by the quality factor, Q, the filling factor, $\eta$, of the microwave-frequency resonator,
and the equilibrium magnetization, $M_0$,
which is determined by the temperature, $T$, and the doping density of $^{31}$P.

In our experiment the static field is $B_0=3500$ Gauss, temperature $T=8$ K, the filling factor
and the quality factor of the resonator are $\eta=0.5$, $Q=500$, respectively.
The doping density of $^{31}$P is very low, $5\times10^{14}$ cm$^{-3}$,
corresponding to a damping time constant $\tau_r\sim 500~\mu$s,
which is much longer than $T_2^*$ and hence the radiation damping effect
is negligible in our DD experiments.

However, for a higher doping density,
of the order of $10^{15{\text -}17}~({\text{cm}})^{-3}$,
RD could be noticeable.
We calculated the influence of RD on the performance of the DD sequence
with $\tau_r=2~\mu$s (corresponding to a doping density $\sim 1.5\times 10^{17}~{\text {cm}}^{-3}$ and other parameters
the same as in the experiments).
We find that in the dynamical decoupling process with imperfect pulses,
RD could induce an asymmetry to the evolution of the initial states
corresponding to the electron spin along the $+z$ and $-z$ directions.

In Sec.~\ref{sec:main_results}, when we studied the effect of pulse errors
on the DD performance,
we neglected the finite pulse duration
and treated the pulses as instantaneous rotations.
RD, however, has its largest impact on spins during the pulses.
If the pulses are ideal, the RD effects that take place
during the delays before and after
each pulse cancel each other.\cite{Abragam}
Moreover, here we focus on the initial states parallel or anti-parallel to $z$-axis.
For a single electron spin during the inter-pulse delays,
the in-plane component of the spin magnetic moment is nonzero
due to pulse errors (but is small).
For a large number of electron spins,
these nonzero components are expected to be evenly distributed in the $x$-$y$ plane,
giving almost zero net in-plane magnetization.
The magnitude of the RD field, ${\bf B}_r$, is proportional to the in-plane magnetization,
hence little RD can be induced.
During the pulse, on the other hand, the in-plane magnetization has nonzero value,
and RD has finite contributions.

We therefore take into consideration the finite duration of the magnetic field pulse.
The pulse is implemented by applying a
driving field to the electron spin for a finite duration, $t_p$.
For the spin state along the $+z$ or $-z$ directions,
at the edges of a pulse
the in-plane magnetization is still small, and the resulting RD is negligible.
This is in contrast with the pulse errors analyzed in Sec.~\ref{sec:main_results},
where contributions from the pulse edges were important.
So here we can assume that the driving field for the pulse is turned on and off abruptly,
and during the pulse the magnitude of the alternating field is constant for each spin.
That is, in the rotating frame the driving field as a function of time is a square wave.
In simulations in the rotating frame
the driving field of the pulse $\pi_{\text X}$($\pi_{\text Y}$) is taken to be along the axis $\vec {\bf n}$($\vec {\bf m}$).
The amplitude of the driving field
is taken as
\begin{equation}
B_p=\bar B_p + \delta B_p
\end{equation}
where the average
$\bar B_p=\pi/(\gamma_e t_p)$
and the deviation
\begin{equation}
\delta B_p=\epsilon/(\gamma_e t_p)
\end{equation}
with $\epsilon$
drawn from distribution Eq.~(\ref{epsdist}).
The parameters for pulse errors (angle errors $\epsilon_x$, $\epsilon_y$ and the rotation axis errors $n_z$, $m_z$)
have the same magnitudes and distributions as specified in Section \ref{sec:err}.
During the pulse, both the static field and the driving field are present.
So in the rotating frame the Hamiltonian, Eq.~(\ref{eq:ham}),
also participates in driving the evolution during the pulse.
We took the magnitude of the driving field $B_{p}=1~\text{Gauss}$ as in experiment
and correspondingly the duration of a $\pi$-pulse is $t_p=0.18~\mu\text{s}$.

We simulated XY CDD with concatenation levels up to 4,
for the electron spin initial states along the $+z$ and $-z$ directions.
The fidelities are denoted as $F_{+z}$ and $F_{-z}$ respectively.
Here $-z$ refers to the equilibrium state of the spin in the static field,
and the $+z$ state is prepared by applying a preparatory $\pi_{\text X}$-pulse
to the spin in the $-z$ state.
Due to the imperfection in the preparatory pulse, the fidelity for $+z$ state
becomes slightly different from the fidelity of the $-z$ state.
See Table~\ref{tb:RD} (A).

\begin{figure}
\includegraphics[angle=270,width=3.5in]{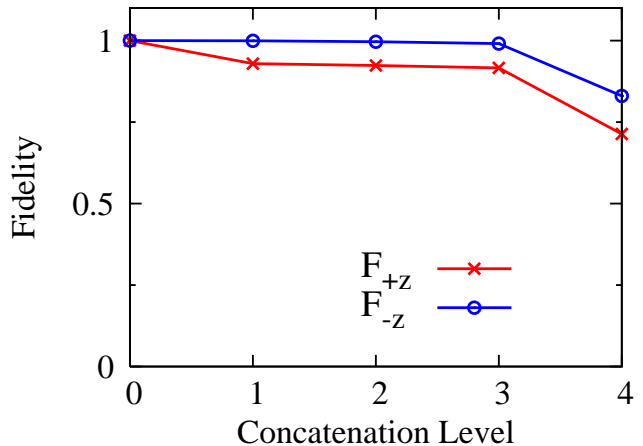}
\caption{\label{fig:cddRD}
Fidelities $F_{+z}$ and $F_{-z}$
for XY CDD sequence with radiation damping considered.
$B_p=1~$Gauss and $\tau_r=2~\mu{\text s}$.
}
\end{figure}

The difference between states $+z$ and $-z$ produced by the preparatory pulse
is further amplified by the RD in the concatenated DD sequence.
Figure~\ref{fig:cddRD} shows the simulation results for XY CDD.
The decoupling protocol preserves the state $-z$ with higher fidelity than $+z$.
The difference is more prominent for concatenation level 4.

This is because RD only takes effect during the magnetic field pulses
when the in-plane magnetization is nonzero.
The RD time constant, $\tau_r$, is larger than the pulse duration by two orders of magnitude.
The effect of RD is therefore very small for the first three concatenation levels, and
after the accumulation during hundreds of pulses,
RD shows up prominently at level 4, (Figure~\ref{fig:cddRD}).

%

To better understand the different roles of the preparatory pulse,
the effect of RD during the pulses and during the inter-pulse delays,
simulations are performed for three cases:
(A) without RD, i.e. $\tau_r=\infty$;
(B) RD is only included during inter-pulse delays;
(C) RD is included throughout the whole evolution
(i.e., both during and between the pulses).
Table~\ref{tb:RD} shows the data for XY CDD in the three cases.
Compared to the results without RD, the difference between $F_{+z}$ and $F_{-z}$
for levels 1-3 is mainly due to the preparatory pulse, and is little affected by RD.
For level-4, RD contributes appreciably.
The different fidelities for level-4 in cases (B) and (C) confirms
that RD contributes mostly during the pulses.

\begin{table}
\caption{\label{tb:RD}
XY CDD pulse sequence. Dependence of the fidelities on the concatenation
level for electron spin initial states along the $+z$ and $-z$ directions.
Pulse field $B_p=1~\text{Gauss}$.
Results for three cases are shown:
(A) $\tau_r=\infty$;
(B) $\tau_r=2~\mu\text{s}$, RD takes effect only during inter-pulse delays.
(C) $\tau_r=2~\mu\text{s}$, RD takes effect all the time.}
\begin{ruledtabular}
\begin{tabular}{ccccccc}
level& (A) $F_{+z}$ & $F_{-z}$ & (B) $F_{+z}$ & $F_{-z}$ & (C) $F_{+z}$ & $F_{-z}$ \\ \hline
$1$ & 0.929 & 0.999 & 0.930 & 0.999 & 0.929& 0.999\\
$2$ & 0.923 & 0.994 & 0.923 & 0.995 & 0.924& 0.993\\
$3$ & 0.928 & 0.998 & 0.926 & 0.998 & 0.920& 0.994\\
$4$ & 0.924 & 0.994 & 0.877 & 0.933 & 0.668& 0.832\\
\end{tabular}
\end{ruledtabular}
\end{table}

Conventionally, starting from the linear Bloch equations,
the states $+z$ and $-z$ would have exactly the same dynamical behavior.
Radiation damping breaks this symmetry.
The RD field, ${\bf B}_r$ (Eq.~(\ref{eq:Br})), is a macroscopic field and
is dependent on the overall
magnetization of the electron spin ensemble.
The Bloch equation is then no longer linear in the magnetization, ${\bf M}$,
resulting in the different dynamics of the quantum states along the $+z$ and $-z$ directions.



\end{document}